\documentclass[12pt]{article}

\usepackage{amsmath}
\usepackage[dvips]{epsfig} 
\usepackage{amssymb}
\def\sech{\rm sech}

\newtheorem{theorem}{Theorem}[section]

\newtheorem{corollary}[theorem]{Corollary}

\begin{document}

\title{Soliton Collisions in the Ion Acoustic Plasma Equations}

\author{Yi Li\thanks{current address: Los Alamos National Laboratories, Los Alamos, New Mexico}\\
University of Minnesota\\ Minneapolis, 55455, USA
\and
D. H. Sattinger\thanks{Research supported by the National Science
Foundation under Grant  DMS-9501233.}\\
University of Minnesota\\Minneapolis 55455, USA}

\maketitle

\centerline{{\it Jour. Mathematical Fluid Mechanics}, {\bf 1}, (1999), 117-130}

\smallskip

{\abstract Numerical experiments involving the interaction of two solitary waves of the ion acoustic plasma equations are described. An exact 2-soliton solution of the relevant KdV equation was fitted to the initial data, and good agreement was maintained throughout the entire interaction. The data demonstrates that the soliton interactions are virtually elastic.}


\section{Introduction}
The ion acoustic plasma equations are
\begin{equation}\label{plasma}
n_t+(nv)_x =0, \qquad
v_t+\left(\frac{v^2}{2}+\varphi\right)_x   = 0, \qquad
\varphi_{xx}-e^\varphi +n  =0,
\end{equation}
where $\varphi$, $n$, and $v$ are respectively the electric potential, ion density and ion velocity.

We wrote a numerical code for these equations and carried out an experiment to see the interaction of two solitary waves. Solitary waves of the plasma equations were obtained numerically. Two solitary waves of different amplitudes were superposed, at a distance sufficiently far apart, and the evolution of the equations with this initial data was computed numerically. 

\smallskip

\centerline{\epsfxsize=\linewidth \epsfbox{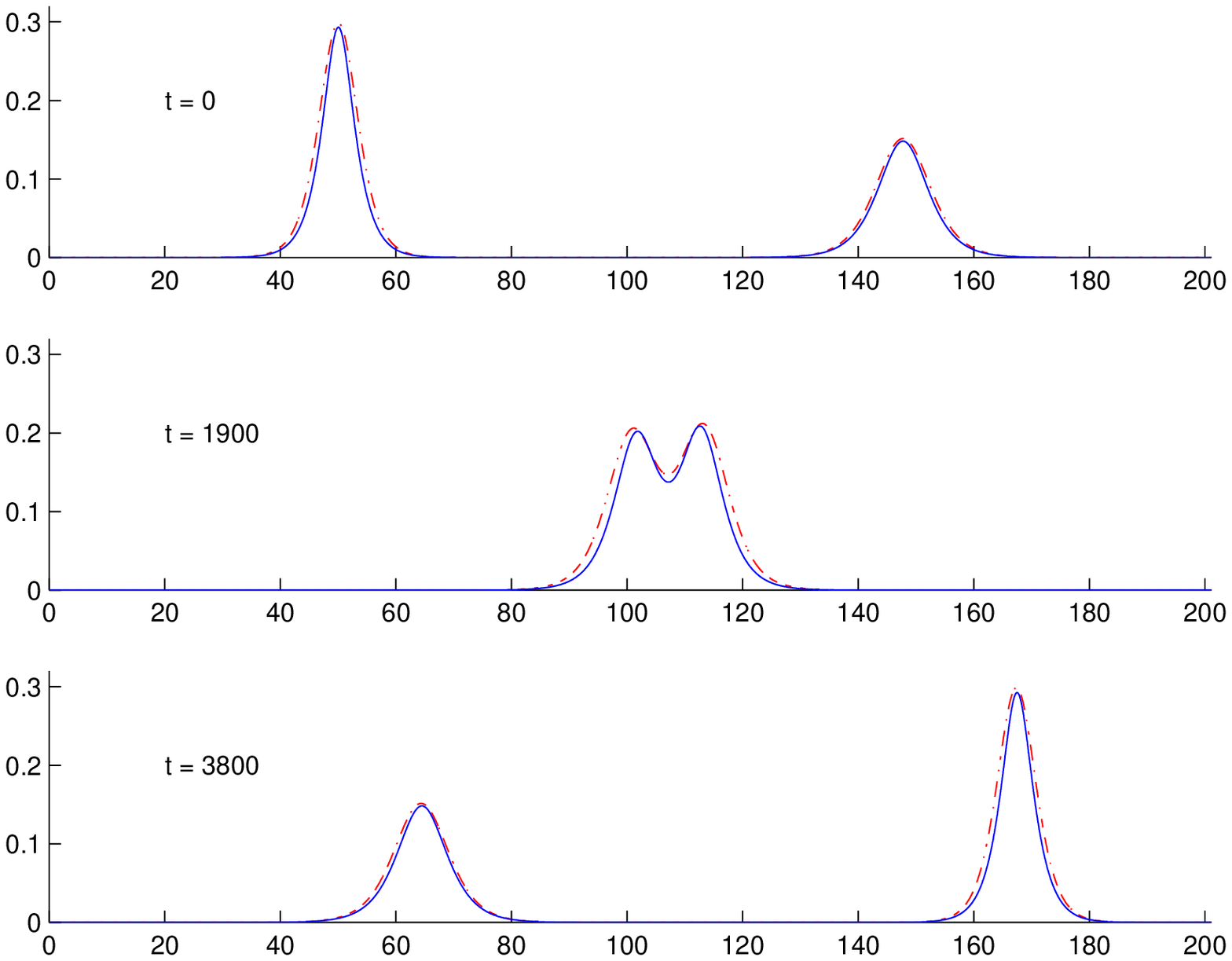}}

\smallskip

\noindent Figure 1: {\small Results of a numerical computation showing the interaction of two solitary waves for the ion acoustic plasma equations, at low amplitudes, by a pseudo-spectral method. The time step was $dt=.008$, $N=2^{13}$ Fourier modes. The computation is done in a reference frame moving with speed 1.07. The speed of the slower wave is 1.05; while that of the larger wave is 1.1.  The solid line is the numerical data, the dashed line an exact KdV 2 soliton solution.}

\smallskip

We fit the numerical data with a 2 soliton solution of the Korteweg-deVries (KdV) equation. Good agreement between the numerical results and the exact two soliton solution is maintained throughout the entire interaction.
While numerical studies of soliton interactions in the literature,  \cite{bps3}, \cite{fenton}, \cite{kodama}, \cite{konno}, \cite{zou}, generally stress the inelastic nature of the interaction, the figures below show that nonelastic effects are negligible in the case of the plasma equations.

\smallskip

\centerline{\epsfxsize=\linewidth \epsfbox{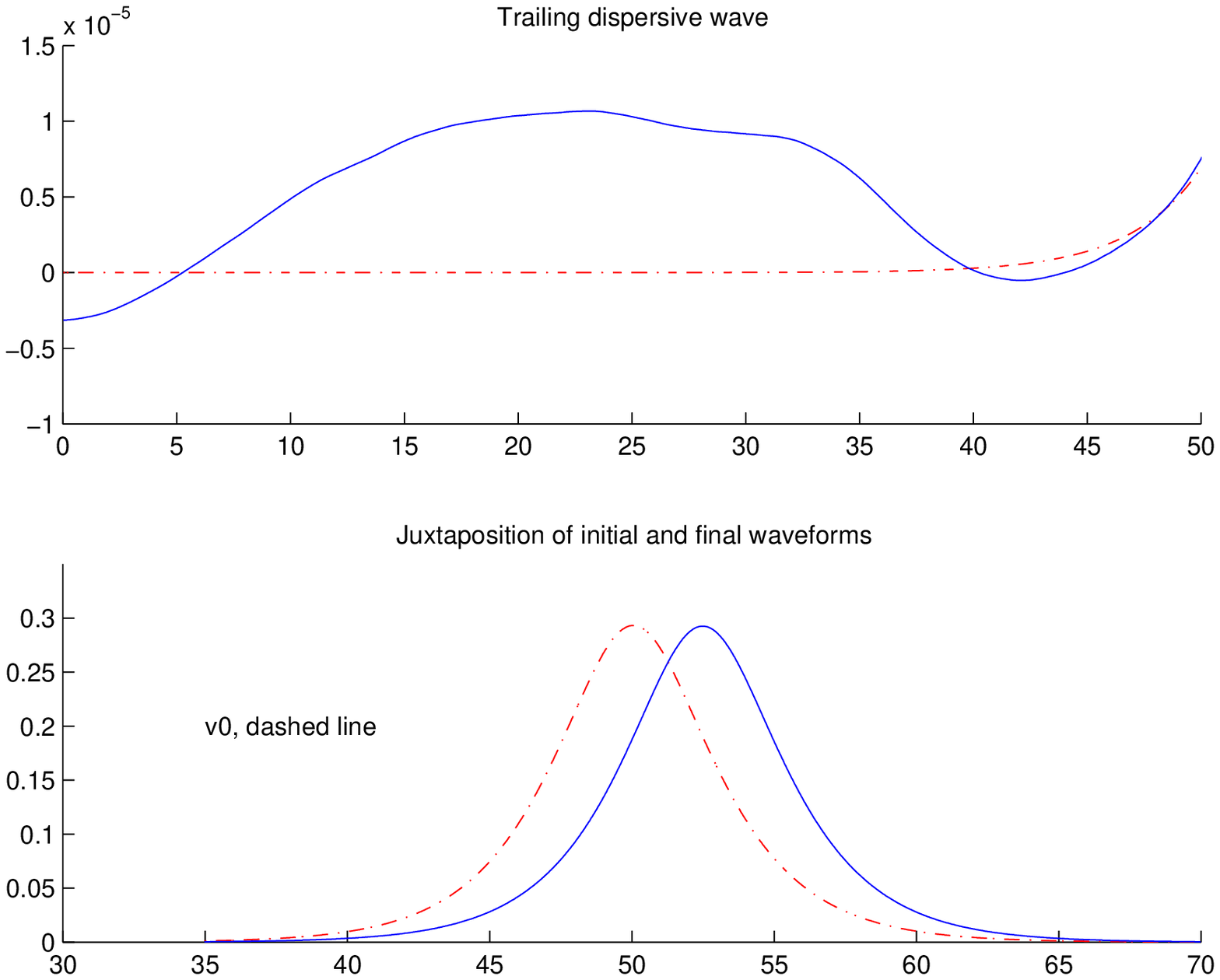}}

\smallskip

\noindent Figure 2: {\small Upper: Trailing dispersive wave at time $t=2600$; the magnitude of the wave is of order $10^{-5}$, compared with an amplitude of .15 of the slower wave. Bottom: If the larger wave is returned to its initial position and superposed on the initial wave form, the results are indisinguishable by graphical comparison; hence we have simply juxtaposed them here. The $H^1$ norm of the difference between the initial waveform and the translation of the final one is
$6.0546\times 10^{-4}$.}

\smallskip

In \S\ref{kdvapprox} we review briefly the singular perturbation argument leading to the KdV approximation, emphasizing the fundamental role of the Galilean and scaling groups. The method for fitting a two soliton solution of the KdV equation to the numerical data using the theoretical formulae for the scattering shifts of the KdV solitons is described in \S\ref{kdvcm}.  As in the Euler equations, there is a wave of maximum speed and amplitude. A comparison of the solitary waves of the KdV equation with that of the full plasma equations is given in \S\ref{sol}.
The numerical methods are discussed in \S\ref{num}.
Finally, in \S\ref{conclusion}, the results of our numerical studies will be compared with the studies in the literature cited above.

\section{The KdV Approximation}\label{kdvapprox}

The Korteweg-deVries (KdV) equation,
arises as a formal singular perturbation of a number of nonlinear dispersive wave equations, such as the Euler equations for an inviscid fluid, and the
ion acoustic equations of plasmas \cite{lmc}, \cite{wash}. Kruskal \cite{kruskal} argues that the KdV equation 
is the unique asymptotically correct model for the Euler or plasma 
equations under these conditions.

Craig \cite{craig} has given a rigorous mathematical proof of the validity of the KdV and Boussinesq approximations
to the Euler equations locally in time.
His results hold for a large class of
initial data, but it is unclear from his analysis whether the validity of the approximation is long enough to cover the interaction of solitary waves.
Apart from this result, we know of no other rigorous results establishing the validity of the KdV approximation. 

The KdV equation is a solvable model. One of its special solutions is the
so-called
``multi-soliton'' solution,

\begin{gather}
u(x,t)=12\frac{d^2}{dx^2}\log \det \Big( \delta_{jk}+ 
\frac{e^{-(\theta_j+\theta_k)}}{\omega_j+\omega_k}\Big), \label{nsol}
\\[4mm]
\theta_j=\omega_j(x-\alpha_j-4\omega_j^2t),
\qquad
0<\omega_1<\dots <\omega_n, \quad \alpha_j\in\mathbb R. \nonumber 
\end{gather}

As is well known, the the solitons interact is elastically: after the interaction, they regain their original shape, and the only evidence of the interaction is a scattering shift. No trailing dispersive waves are generated.

The Euler equations themselves are complicated, being a free boundary problem. 
The plasma equations are considerably simpler to deal with, both numerically and analytically, and are also an interesting physical model; hence, in order to investigate the validity of the KdV approximation, it makes sense to consider the plasma
equations themselves. 

The plasma equations
are Galilean invariant. This means that the equations are the same in any Galilean frame, and we can 
shift to a moving frame of reference simply by subtracting the speed of the moving frame from the velocity $v$. In the moving frame, the velocity $v'$ tends to $-c$ at infinity. This amounts to expanding about the quiescent state $n=1,v=-c,\varphi=0$, 
where $c$ is the velocity of the reference frame.

In particular, $v=-1$ at infinity in a Galilean frame moving with speed 1, and the dispersion relation in this reference frame is found to be
$$
\omega=-k+\frac{k}{\sqrt{1+k^2}}=k(-1+[1-\frac12 k^2+\dots]\doteq -\frac12 k^3,
$$
in the long wave approximation.

The natural scaling associated with this long wave approximation is $x'=\varepsilon x$, $t'= \varepsilon^3 t$. Introducing
this scaling into  \eqref{plasma}, we obtain (after division by $\varepsilon$)
\begin{eqnarray*}
\varepsilon^2n_{t'}+(nv)_{x'} =0, \quad
\varepsilon^2 v_{t'}+\left(\frac{v^2}{2}+\varphi\right)_{x'}  = 0, \quad
-\varepsilon^2\varphi_{x'}+e^\varphi = n. 
\end{eqnarray*}

This perturbation scheme is singular, since the character of the equations is
changed when $\varepsilon=0$.
Since only $\varepsilon^2$ appears in these equations, we formally expand all quantities in powers
of $\varepsilon^2$ about $n=1,v=-1,\varphi=0$.
When we do this, substitute the expansions into the above equations, and collect terms,
we find at first order that $n_1=v_1=\varphi_1.$

At next order we obtain
\begin{align*}
n_{1,t'}+(n_1v_1)_{x'} +(v_2-n_2)_{x'} &=0, \qquad
v_{1,t'}+\left(\frac{v_1^2}{2}-v_2+\varphi_2\right)_{x'}  = 0, \\[4mm]
-\varphi_{1,x'x'}+\varphi_2+\frac12 \varphi_1^2 &= n_2. 
\end{align*}
The second order quantities $n_2,\,v_2,$ and $\varphi_2$ may be eliminated from this system; and, dropping the primes, we obtain
the Korteweg-de Vries equation for $v_1$:
\begin{equation}\label{kdv}
v_{1,t}+v_1v_{1,x}+\frac12 v_{1,xxx}=0.
\end{equation}

\section{Comparison with KdV}\label{kdvcm}

In order to compare the numerical data with solutions of the KdV equation, it is important to point out that one does not simply solve the KdV equation with the initial data from the plasma equations. This would produce very bad results, for two reasons. First, the plasma solitary wave deviates from the KdV solitary wave, as we show in \S \ref{sol}; so that if this data were taken as initial data for the KdV equation, secondary wave trains would develop, and non-elastic behavior would be observed. Moreover, there is a shift in the amplitudes and phases due to the soliton interaction, \cite{zou}. We now discuss the method of fitting an exact two soliton solution of the Korteweg deVries equation to the numerical data.

First note that there is a factor of 1/2 in the KdV approximation \eqref{kdv}. This is easily accounted for by a simple rescaling:
$$
v(x,t)=\frac12 u\left(x,\frac12 t\right),
$$
where $u$ satisfies the KdV equation $u_t+uu_x+u_{xxx}=0.$

By expanding the determinant in \eqref{nsol}, and renaming the phase shifts $\alpha_1$, $\alpha_2$, the two-soliton solution of can be written
$$
v=6\frac{d^2}{dx^2}\log\tau(\theta_1,\theta_2), \quad \tau=1+e^{-2\theta_1}+e^{-2\theta_2}+e^{-2(\theta_1+\theta_2+\alpha)},
$$
where
\begin{eqnarray*}
\theta_j=\omega_j(x-\alpha_j-(c+2\omega_j^2)t), \ \  j=1,2;
\qquad
\alpha=\log\frac{\omega_2+\omega_1}{\omega_2-\omega_1}.
\end{eqnarray*}

The two soliton solutions form a four parameter family, $\alpha_1,\alpha_2,\omega_1,\omega_2$. The linearized KdV equation at the 2 soliton solution therefore has a four dimensional null space, obtained by differentiating the equation with respect to the four parameters. 
In the perturbation series, the four parameters $\alpha_1,\alpha_2,\omega_1,\omega_2$ must be
adjusted in order to eliminate the resonance terms which lie in the null space of the linearized operator, as discussed by Zou and Su \cite{zou}, who calculate the shift in these four parameters at second and third order and compare them with the numerical data. 

In the present case, the shifts in $\alpha_1,\delta \omega_2$ may be determined from the numerical data as follows.
The locations of the large and small waves before and after the interaction lead to four equations in four unknowns. Moreover, the waves are sufficiently separated before and after the interaction that we can apply the known formulae for the phase shifts incurred in the interaction \cite{nmpz}.

The two soliton solution has the asymptotic behavior (cf. \cite{hs})
$$
u=6\frac{d^2}{dx^2}\log\tau \sim 6\omega_1^2\sech^2(\theta_1+\alpha) +6\omega_2^2\sech^2\theta_2,
\qquad
\to\infty;
$$
and from the knowledge of the scattering shifts we find
that
$$
u\sim 6\omega_1^2\sech^2\theta_1+6\omega_2^2\sech^2(\theta_2+\alpha),
\qquad
t\to -\infty.
$$

Therefore, our matching conditions give
\begin{gather*}
t=0: \qquad \theta_1=0, \qquad
\theta_2+\alpha=0;\\[4mm]
t=T: \qquad \theta_1+\alpha=0, \qquad \theta_2=0.
\end{gather*}
These four conditions lead to the equations
\begin{eqnarray*}
\alpha_1=x_1^-=x_1^+-(c+2\omega_1^2)T+\frac{1}{\omega_1}\log\frac{\omega_2+\omega_1}{\omega_2-\omega_1},\\
\alpha_2=x_2^+-(c+2\omega_1^2)T=x_2^-+\frac{1}{\omega_2}\log\frac{\omega_2+\omega_1}{\omega_2-\omega_1}.
\end{eqnarray*}
Here $x_j^\pm$ denote the locations of the $j^{th}$ wave at times $t=0$ and $t=T$, the total elapsed time, and $c$ is the relative speed of the computation frame to the Galilean frame. 

The phase constants can be eliminated from these equations, and we obtain two equations in $\omega_1$ and $\omega_2$:
\begin{eqnarray*}
2\omega_j^2T+(-1)^j\frac{1}{\omega_j}\log\frac{\omega_2+\omega_1}{\omega_2-\omega_1}=\delta x_j-cT\qquad j=1,2,
\end{eqnarray*}
where $\delta x_j$ is the total distance traversed by the $j^{th}$ wave.

We solved the equations for $\omega_1$, $\omega_2$ iteratively. 
As an initial guess we took the values obtained by matching the speeds of the two KdV waves exactly with the speeds of the soliton waves. The speed of the solitary wave $6\omega^2\sech^2(\omega(x-2\omega^2t)$ is $2\omega^2$. The speeds of the two solitary plasma waves (relative to the Galilean frame) are .05 and .1. Therefore, as a first approximation, we take
$\omega_1=\sqrt{.05/2}=.1581$ and $\omega_2=\sqrt{.1/2}=.2236.$

Our data are 
\begin{gather*}
T=3800; \qquad c=-.07; \qquad x_1^-=147.7349; \qquad x_2^-=50.0468;\\[4mm]
\delta x_1=-85.2439; \qquad
\delta x_2=120.4166
\end{gather*}
We obtained $\omega_1=.1589$ and $\omega_2=.2231.$
 The relative phases $\alpha_1$ and $\alpha_2$ are then determined by any of the four equations above.

\section{Solitary waves}\label{sol}

The plasma equations model two-directional propagation of ion acoustic waves,
hence the system possesses solitary waves travelling in either
direction. 
Working in a frame moving with the wave, we obtain \cite{sagdeev}
\begin{eqnarray}
n=\frac{v}{c-v}, \qquad v=c-\sqrt{c^2-2\varphi} \label{3}\\
\varphi''=f(\varphi), \qquad f(\varphi)=e^\varphi-\frac{c}{\sqrt{c^2-2\varphi}}\label{4}
\end{eqnarray}

A straightforward analysis of this second order equation yields the following facts,
which we summarize here.
 Equation \eqref{4} has no solitary wave solution
 when $0<c<1$ since $ f $ is negative in a neighborhood of the origin; 
however, the origin is a stable center, and the
plasma equations 
 have a two parameter family (amplitude and phase) 
of periodic waves for all $0<c<1$.

When $ c>1 $, the origin is a saddle point. Since $ f'(\varphi)< 0 $ for all $ \varphi<0$, there are no homoclinic orbits in the left half phase plane. In the following theorem, we give a necessary and sufficient condition
 for the existence of a homoclinic orbit, which physically is a solitary wave of elevation.
 
\begin{theorem}\label{thm1} Let $ \bar c$  ( $ \approx 1.5852 $ ) be the positive root of the equation
 $$ 
e^{c^2/2 }=1+c^2.   
$$
 For each $ c\in (1, \bar c) $, there is a unique solitary wave of
 elevation  to equation \eqref{4} 
 which is even and real-analytic. When $ c=\bar c$,
 \eqref{4} also has a continuous solitary wave solution with unbounded second derivative at the origin, which is a weak solution of \eqref{4} in the sense of distributions.  
 \end{theorem}  
 We omit the proof, which is not difficult.

 \smallskip
\centerline{\epsfxsize=\linewidth \epsfbox{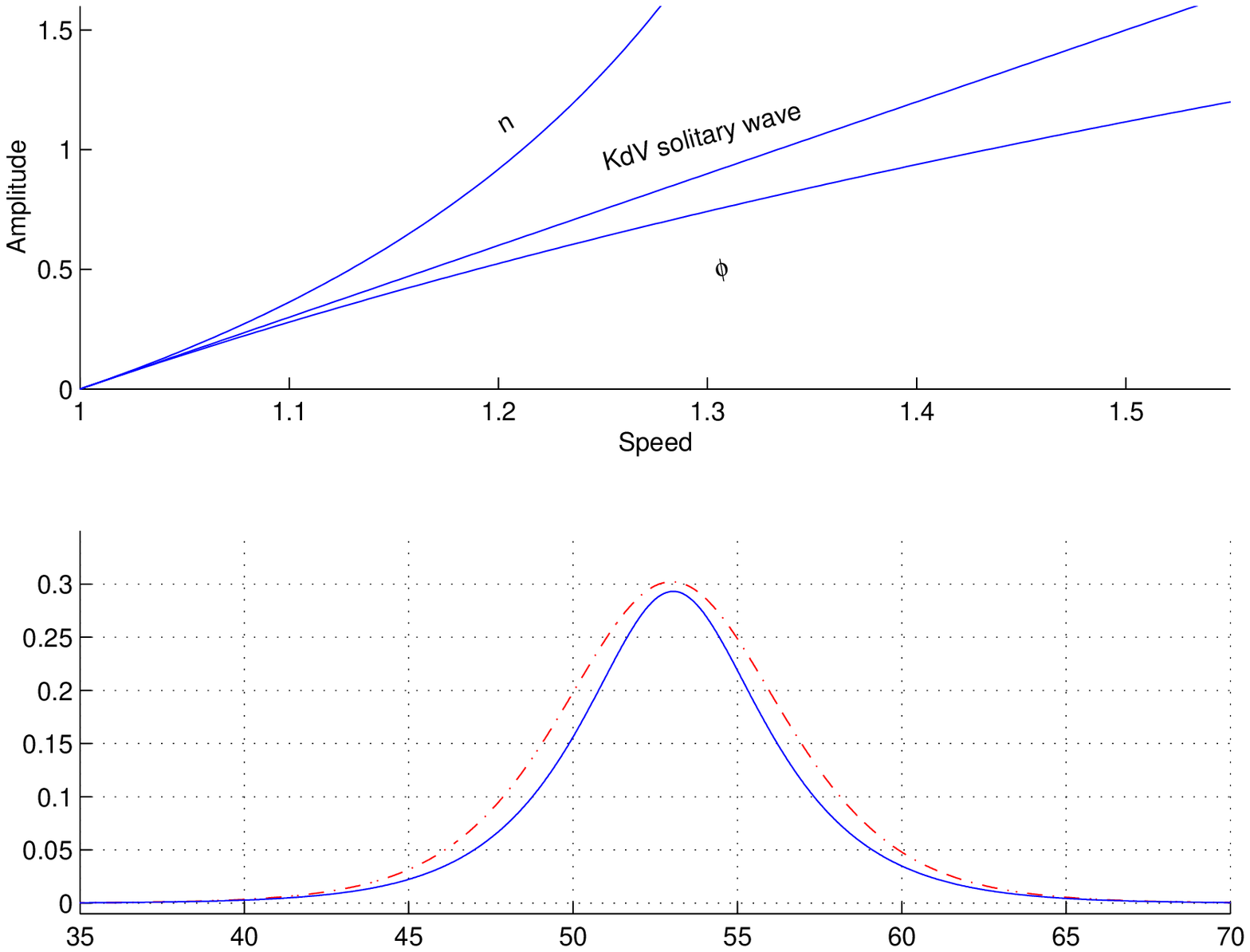}}

\smallskip

\noindent Figure 3. {\small Upper: The relationship of amplitude to speed for the $n$ and $\varphi$ components of the plasma wave compared with that for the KdV solitary wave, which is linear.
The amplitude of $n$ blows up as $c\to\bar c$, the maximum wave speed.
Lower: Comparison of the single solitary KdV solution $6\omega^2\sech^2\omega x$, (dashed), to the $\varphi$ wave of the plasma equations (solid).}

\smallskip

\begin{theorem}\label{thm2} For each constant 
$c$ such that $1<c<\bar c$, there is a unique solitary wave solution of \eqref{plasma} the form
$$
\{1+n(x-ct),\,v(x-ct),\,\varphi(x-ct)\}
$$
where $n,v,\varphi$ tend to zero at infinity. When $c=\bar c$ the solution is a weak
solution in the sense of distributions.

Left traveling waves are obtained by the reflection symmetry $c\mapsto -c$,
$v\mapsto -v$.
 \end{theorem}

We can always make a Galilean transformation so that the asymptotic velocity at infinity is zero. A direct consequence of Theorems \ref{thm1} and \ref{thm2} is the convergence of periodic
 orbits to solitary wave solutions. 

\begin{corollary}\label{cor3}  Each solitary wave solution of the plasma equations
  is a limit of a sequence of periodic waves.
 \end{corollary}

\section{The numerical scheme}\label{num}

In this section we describe briefly the implicit pseudo-spectral scheme \cite{wineberg} we used for numerically solving the ion plasma equations. 
We write the plasma equations in a reference frame moving with speed $c$; we also replace $n$ by $n+1$, so that all three variables tend to zero at infinity. Note that this is not a 
Galilean transformation, since the velocity $v$ still vanishes at infinity, and the form of the
equations is changed. We then get 
\begin{align*}
n_t-cn_x+v_x+(nv)_x=&0, \\[4mm]
v_t-cv_x+\left(\frac{v^2}{2}+\varphi\right)_x=&0,\\[4mm]
e^\varphi-1-\varphi+(1-D^2)\varphi=&n
\qquad D=d/dx.
\end{align*}

We integrate the first two equations with respect to time over the interval $(t_1,t_2)$, where $t_2=t_1+\delta t$. Denote $n_1=n(t_1)$ and $n_2=n(t_2)$. We get, in the first equation,
$$
n_2-n_1+\int_{t_1}^{t_2} D(v-cn+nv)\,dt=0.
$$
We approximate the integral using the trapezoid rule, obtaining
$$
n_2-n_1+\frac12\delta t D(v_2-cn_2+v_1-cn_1)+\frac12\delta t D(n_2v_2+n_1v_1)=0.
$$
The equation in $v$ is treated in the same way, and we obtain there
$$
v_2-v_1-\frac{c}{2}\delta t D(v_1+v_2)+\frac{\delta t}{2}D\left(\frac{v_1^2}{2}+\varphi_1+\frac{v_2^2}{2}+\varphi_2\right)=0.
$$
We combine these equations together with the equation for $n_2$ and $\varphi_2$ into a $3\times 3$ nonlinear system, which we write in the form
$$
\begin{pmatrix} A& B & 0 \\[4mm]
0 & A& B\\[4mm]
-1 & 0 & C \end{pmatrix} \begin{pmatrix} n_2 \\[4mm] v_2 \\[4mm] \varphi_2 \end{pmatrix} \\[4mm]
=\begin{pmatrix} A_1& -B& 0 \\[4mm]
0 & A_1& -B\\[4mm]
0& 0 & 0 \end{pmatrix} \begin{pmatrix} n_1 \\[4mm] v_1 \\[4mm] \varphi_1 \end{pmatrix} 
+\begin{pmatrix}  -B(n_1v_1+n_2v_2) \\[4mm]
-B\left(\frac{v_1^2}{2}+\frac{v_2^2}{2}\right)\\[4mm]
K(\varphi_1)+K(\varphi_2)\end{pmatrix},
$$
where
$$
A=1-\frac{c\delta t}{2}D ,
\qquad
A_1=1+\frac{c\delta t}{2}D ,
\qquad
B=\frac{\delta t}{2}D ,
\qquad
C=1-D^2,
$$
and
$$
K(\varphi)=\frac12 (1+\varphi-e^\varphi).
$$
We found by experimentation that averaging the nonlinear term in the third equation produced slightly more accurate results when integrating the solitary wave.

Inverting the $3\times 3$ matrix on the left we obtain a nonlinear implicit scheme for the time step. This is solved iteratively. The size of the time step is constrained by the requirement that the iteration scheme converge sufficiently rapidly.

The equations are integrated in a frame moving with speed 1.07. The speed of the larger wave is 1.1; that of the smaller wave is 1.05. The speed of the maximum wave is approximately 1.5852.
In the moving frame, the smaller wave drops back, while the larger wave advances. This has the advantage of keeping both waves within a fixed interval throughout the interaction.

The numerical computations are complicated by small 'discontinuities' at the endpoints of the interval. Though the solitary wave decays exponentially fast, and is of the order of $10^{-6}$ at the endpoints of the interval, there is nevertheless a small jump at the endpoints, if one uses the homoclinic solution, rather than a periodic wave nearby. Therefore, rather than computing the solitary wave, which is a homoclinic orbit, we instead computed a periodic wave very close to the homoclinic orbit.

Numerical error introduces high frequency noise, which is transferred to higher modes by the nonlinear terms. For the KdV equation this effect causes no problems in the numerical computations, since the dispersion relation of the KdV equation grows like $k^3$, and acts like a high frequency cut-off in the implicit pseudo-spectral scheme.

But the plasma equations have much weaker dispersion, 
and high frequency energies are not attenuated, with the result that the the calculations are corrupted over time. Because of the scaling involved, the code must be integrated over a very long time scale in order to see the interaction of the two solitary waves, in this case, to T=2600. 

To compensate for this problem, we periodically filtered the data using 
the `sharpened raised cosine' filter (cf. \cite{can}, p. 248):
$$
\sigma(\theta)=\sigma_0^4(35-84\sigma_0+70\sigma_o^2-20\sigma_0^3),
\qquad
\sigma_0=\frac12 (1+\cos\theta).
$$
The initial data was filtered; and the solution was filtered at time intervals of 50 units. 

The filtering, of course, removes energy from the system; but we calculated the energy
$$
{\cal E}=\int n(\varphi+cv+\frac12 v^2)-e^\varphi-\frac{\varphi_x^2}{2}\,dx
$$
over the course of the interaction. It deviated from the original values by 
 , showing that the filtering is relatively mild.

We tested the numerical scheme by numerically intergrating a single solitary wave in a reference frame moving with the wave. The calculations indicate 
the solitary wave to be fixed  point of the numerical scheme. 

\section{Conclusions}\label{conclusion}

We draw two main conclusions from our study: 1) the solitary wave interactions are virtually elastic; 2) the KdV equation itself is qualitatively accurate in modelling the elastic interaction of solitary waves. 

Numerical computations on the second and third order terms in the 
singular perturbation expansions for the Euler equations were carried 
out by Zou and Su \cite{zou}.  They find that the collision is elastic 
up to second order; but the computation of the third order term shows 
``a secondary wave train trailing behind the smaller wave after the 
collision.''
They state that 

\smallskip

``The third-order solution contains a secondary wave train. This result may be considered to be consistent with the suggestion given by Benjamin and Olver, (\cite{bo}) and later Olver (\cite{o}), that perfect interactions of solitary waves akin to those of KdV solitons are unlikely for the complete water wave problem. . . ''
\smallskip

There is indeed a dispersive tail behind the smaller wave after the interaction, as indicated in Figure 2; but it is of extremely small amplitude: approximately $10^{-5}$, compared with the amplitude of the smaller wave, which is approximately .15. Hence the relative amplitude of the trailing wave is roughly $.6667\times 10^{-4}$.  The small parameter in our theory is $\omega\approx .2$; hence third order terms are of order $\omega^6=.64\times 10^{-4}$. This is consistent with the dispersive wave at third order found by Zou and  Su.

A numerical study of the first and second order terms in the KdV 
approximation for the ion acoustic plasma equations was carried out 
by Konno {\it et.al.} \cite{konno}. They found secondary waves at second 
order; but they did not cast out the secular terms in their expansion, 
as did Zou and Su.

An alternative small amplitude model, for the Euler equations has been proposed by Benjamin \it et.al. \rm \cite{bbm}, and is now known as the 
regularized long wave, or BBM equation.  
  Numerical experiments by Bona, Pritchard and Scott \cite{bps3} on the two soliton interaction of the BBM equation show a somewhat larger dispersive tail of relative magnitude $10^{-2}$. Their results are confirmed by Kodama \cite{kodama}, who also shows that inelastic interactions can be expected in dispersive systems which are not completely integrable.

A numerical investigation of the two soliton interaction of the full Euler equations was carried out by Fenton and Rienecker \cite{fenton}. They found good agreement with the KdV approximation, although `some deviations from the theoretical predictions were observed--the overtaking high wave grew significantly at the expense of the low wave, and the predicted phase shift was found to be only roughly described by the theory.' Comparing their results to the BBM equation, they state (p. 427) that their computations `showed {\it no trace} of trailing waves, either from plots of the free surface or, more convincingly, by second differences of the point values of surface elevation'. 

Further investigation of the soliton interactions in the Euler equations seems warranted. Our results for the plasma equations show an excellent  fit to the theoretical predictions of the KdV equation. Whether this is due to fundamental differences between the Euler and plasma equations, or due to the fact that the Euler equations are more difficult to integrate, is not clear.  The Euler equations require an accurate approximation to the Dirichlet to Neumann map on the free surface, an issue which was not explicitly discussed in \cite{fenton}.

In our view, the inelasticity of the interaction has received undue emphasis in the literature. This is misleading, since it would suggest that the KdV equation is not a valid model. The inelastic nature of the soliton interactions at low amplitudes are negligible, and the KdV equation models
the interaction of overtaking solitary waves with good accuracy. 

We calculated the relative error of the KdV approximation to the $v$ wave of the plasma data in the $H^1$ norm. We calculated the $H^1$ norm of the difference between the KdV solution and the $v$ wave, and divided by $||v||_{H^1}$. The ratio ranged from approximately .15 at $t=0$ to a maximum of .2 during the interaction, then dropped back to a range of .15. This is relatively high, indicating that we are operating somewhat away from the KdV approximation. This is clear from Figure 3, which shows a comparison of the KdV solitary wave with $v$. 

The qualitative prediction of the KdV model of elastic interactions of solitary waves is substantially correct. The discrepancies between the KdV equation and the full equations show that higher order terms in the singular perturbation expansion must be taken into account in order to obtain a quantitative agreement.  If anything, these two facts indicate that the elasticity of the interaction is robust, in that the interaction is virtually elastic even though there is a relatively large error in the KdV approximation itself.

One natural question, considered by Moser and Sachs, \cite{sachs}
is the following:
Do the multi-soliton solutions of the KdV equation extend to
solutions of the Euler equations on $0\le t<\infty$? A positive answer to 
this question would demonstrate
the global existence of a non-trivial, time dependent solution of the Euler
equations,
something which has not yet been done. 

Moser's conjecture would be established by proving the convergence of
the singular perturbation approximation on $0<t<\infty$.
This would also 1) establish the validity of the KdV approximation on an infinite time scale; 2) establish uniform estimate in time on the error term.

Sachs \cite{sachs} obtained a representation of the propagator for the linearized KdV equation. Using his representation, Haragus and Sattinger \cite{hs} derived uniform estimates for the propagator of the linearized KdV equations about an $n$-soliton solution on $\mathbb R\times\mathbb R_+$. The linearized equations have a $2n$ dimensional null space; and a Fredholm alternative for the time dependent operator in an appropriate function space was proved.

In the case of the Euler and plasma equations, there is no general existence theorem, nor an $H^1$ estimate, as in the case of Pego and Weinstein's proof \cite{pw1} of the stability of the solitary waves of the generalized KdV equation; hence their analysis does not directly apply here. We anticipate that a Nash-Moser type of implicit
function theorem \cite{moser}, \cite{scheurle} may be needed.
A natural space to work in is the space of functions analytic in a
strip
in the complex $x$ plane. Estimates for the linearized operator in this class were obtained in \cite{hs} which were uniform on $0<t<\infty$. The implications of this are that the perturbation series itself does not converge but is only asymptotically valid. The actual solution would be $C^\infty$, and accurate to all orders, but not analytic in $\varepsilon$.

\end{document}